\input harvmac
\overfullrule=0pt

%
\def\sqr#1#2{{\vbox{\hrule height.#2pt\hbox{\vrule width
.#2pt height#1pt \kern#1pt\vrule width.#2pt}\hrule height.#2pt}}}
\def\Box{\mathchoice\sqr64\sqr64\sqr{4.2}3\sqr33}
\def\t{{\theta}}
\def\s{{\sigma}}

\def\half{{\textstyle{1\over 2}}}

\def\p{{\partial}}

\def\half{{\textstyle{1\over 2}}}

\def\thalf{{\textstyle{3\over 2}}}

\def\t{{\theta}}

\Title{ \vbox{\baselineskip12pt
\hbox{hep-th/0201027}}}
{\vbox{\centerline{Three-Graviton Amplitude in }
\bigskip
\centerline{Berkovits-Vafa-Witten Variables}}}
\smallskip
\centerline{K. Bobkov and L. Dolan
}
\smallskip
\centerline{\it Department of Physics}
\centerline{\it
University of North Carolina, Chapel Hill, NC 27599-3255}
\bigskip
\smallskip

\noindent
We compute the three-graviton tree amplitude in Type IIB superstring
theory compactified to six dimensions using the manifestly (6d) supersymmetric
Berkovits-Vafa-Witten worldsheet variables. We consider
two cases of background geometry: the flat space 
example ${\bf R^6}\times K3$,
and the curved example 
$AdS_3\times {\bf S^3}\times K3$ with Ramond flux, and
compute the correlation functions in the bulk.
\Date{}

\nref\bv{N. Berkovits and C. Vafa, ``$N=4$ Topological Strings'',
Nucl. Phys. B433 (1995) 123, hep-th/9407190.}

\nref\be{N. Berkovits, ``A New Description of the Superstring'',
Jorge Swieca Summer School 1995, p. 490, hep-th/9604123.}


\nref\bvw{N. Berkovits, C. Vafa, and E. Witten, ``Conformal Field Theory
of AdS Background With Ramond-Ramond Flux'', JHEP 9903: 018, 1999;
hep-th/9902098.}

\nref\dw{L. Dolan, and  E. Witten, ``Vertex Operators for
AdS3 Background With Ramond-Ramond Flux'', JHEP 9911:003, 1999;
hep-th/9910205}

\nref\r{L. Romans, ``Self-Duality for Interacting Fields'',
Nucl. Phys. B276 (1986) 71.}

\nref\sez{S. Deger, A. Kaya, E. Sezgin and P. Sundell,
``Spectrum of $D=6, N=4b$ Supergravity on $AdS_3 \times S^3$'',
Nucl. Phys. {\bf B536} (1988) 110; hepth/9804166.}

\nref\btenflat{N. Berkovits, ``Super-Poincare Covariant Quantization
of the Superstring'', JHEP 0004: 018, 2000; hep-th/0001035.}

\nref\btenflatamp{N. Berkovits, B.C. Vallilo, ``Consistency of 
Super-Poincare Covariant Superstring Tree Amplitudes'',
JHEP 0007: 015, 2000; hep-th/0004171.}

\nref\bcoh{N. Berkovits, ``Cohomology in the Pure Spinor
Formalism for the Superstring'', JHEP 0009: 046, 2000;
hep-th/0006003.}

\nref\bannarb{N. Berkovits, ``Covariant Quantization 
of the Superstring'', Presented at Strings '00, Ann Arbor, Michigan,
Int. J. Mod. Phys. A16:801, 2001; hep-th/0008145.}

\nref\bvoadsfive{N. Berkovits and O. Chandia, 
``Superstring Vertex Operators in an $AdS_5\times S^5$ Background'',
Nucl. Phys. B596: 185, 2001; hep-th/0009168.}

\nref\brns{N. Berkovits, ``Relating the RNS and Pure Spinor Formalisms
for the Superstring'', JHEP 0108: 026, 2001;
hep-th/0104247.} 

\nref\blicoh{N. Berkovits and O. Chandia,
``Lorentz Invariance of the Pure Spinor BRST Cohomology for the
Superstring'', Phys. Lett. B514: 394, 2001;
hep-th/0105149.}

\nref\bpz{ A. Belavin, A. Polyakov, and A. Zamolodchikov,
``Infinite Conformal Symmetry in Two-Dimensional Quantum Field Theory'',
Nucl. Phys. {\bf B241} (1984) 333.}

\nref\fms{D. Friedan, E. Martinec, and S. Shenker, ``Conformal Invariance,
Supersymmetry and String Theory'',
Nucl. Phys. {\bf B271} (1986) 93.}

\nref\lt{D. Lust and S. Theisen, {\it Lectures on String Theory}
Lecture Notes in Physics vol. 346 (Springer-Verlag, 1989).} 

\nref\cf{J. Cohn, D. Friedan, Z. Qiu, and S. Shenker, ``Covariant
Quantization of Supersymmetric String Theories: the Spinor Field 
of the Ramond-Neveu-Schwarz Model'', 
Nucl. Phys. {\bf B278} (1986) 577.}

\nref\rs{N. Read and H. Saleur, ``Exact Spectra of Conformal Supersymmetric
Nonlinear Sigma Models in Two Dimensions'', Nucl. Phys. B613: 409, 2001;
hep-th/0106124.}

\newsec{Introduction}

We compute string tree correlation functions 
using the manifestly supersymmetric
covariant formulation of Berkovits-Vafa-Witten (BVW) for Type IIB superstrings
compactified to six dimensions \refs{\bv-\sez}. Unlike the 
ten-dimensional covariant pure spinor quantization \refs{\btenflat-\blicoh}, 
the six-dimensional version\refs{\bv-\bvw}
we use here incorporates an $N=4$ topological
string formulation to compute tree level scattering amplitudes.

For IIB superstrings either on flat six-dimensional space times $K3$,
or on $AdS_3\times{\bf S^3}\times K3$,
the massless degrees of freedom correspond to  
a $D=6$, $N=(2,0)$ supergravity multiplet and 21 tensor multiplets.
In this paper, we consider the string theory tree level scattering
of three gravitons both in the flat case and the $AdS_3\times S^3$
case, where the latter has background Ramond flux.
For these amplitudes, 
the relevant massless compactification independent vertex operator, in the
BVW worldsheet formalism, contains the graviton, dilaton and two-form field
which contribute to the supergravity and one of the tensor multiplets.

In this formalism \refs{\bv-\bvw}, 
the Type IIB superstring compactified on $K3$, which has 16 supercharges
corresponding to 16 unbroken supersymmetries, 
has 8 which are manifest in that they act geometrically on the target
space. These are given by $F_a,\bar F_a$ described below and are related
to the presence of 8 theta world sheet
variables $\theta^a,\bar\theta^a$.
The other 8 supersymmetries $E_a,\bar E_a$
are not manifest, but can still be expressed in 
terms of ordinary world sheet fields, {\it i.e.} not
spin operators. Therefore, in addition 
to making some of the supersymmetry geometric (either in
${\bf R^6}$ or $AdS_3\times {\bf S^3}$), 
this formalism is also advantageous to 
describing background fields belonging to the Ramond-Ramond (RR) sector,
since the worldsheet fields which couple to the RR background fields
are not spin fields. Thus for strings on AdS that require RR backgrounds,
the purpose of using BVW variables is that
RR background fields can be added to the worldsheet action without 
adding spin fields to the worldsheet action \refs{\bvw}. 

On flat space (${\bf R^6}$), the BVW variables describe a free
worldsheet conformal field theory. Their operator products (OPE's)
are reviewed in sect. 2, together with the $N=4$ topological
method for computing string correlation functions. In sect. 3
we use these OPE's to evaluate the flat space three graviton tree amplitude
in position space, and show that it reduces to the conventional answer. 

On curved space, the BVW variables do not satisfy free operator
product relations. Nonetheless, we proceed to evaluate the
three graviton correlation function on $AdS_3\times {\bf S^3}$
by assuming a form for the OPE's. It is motivated by \refs{\dw},
where the vertex operator constraint equations were derived
for $AdS_3\times {\bf S^3}$ by requiring them to be invariant 
under the $AdS$ supersymmetry transformations.
It can be shown that our assumed OPE's result in
the same constraint equations. In sect. 4, we compute 
the curved space three gravition tree amplitude using these OPE's.

\newsec{Review of Components}

The $N=4$ topological prescription \refs{\bv-\bvw}\ for calculating 
superstring tree-level amplitudes is
\eqn\harry{<\,V_1(z_1)\,(\tilde G^+_0 V_2(z_2))
(G^+_0 V_3(z_3)) \prod_{r=4}^n\int dz_r G^-_{-1} G^+_0 V(z_r)\,>\,.}
where $G_n^\pm$ are elements of the topological $N=4$ super Virasoro
algebra, and the notation  $O_n V(z)$ denotes the pole of order $d+n$ in the
OPE of $O(\zeta)$ with $V(z)$, when
$O$ is an operator of conformal dimension $d$.

\subsec{$N=4$ Superconformal Algebra in Flat Space}

For the IIB superstring there is both a holomorphic $N=4$ superconformal
algebra and another anti-holomorphic one. The holomorphic 
generators, specialized for the IIB string compactified to six dimensions,
and in terms of BVW worldsheet variables, are \refs{\bvw}
\eqn\svgen{
\eqalign{\tilde T (z) &
= -\half\partial x^m\partial x_m - p_a \partial\theta^a
-\half\partial\rho\partial\rho - \half\partial\sigma\partial\sigma
+ \thalf\partial^2 (\rho + i\sigma) + \tilde T_C\cr
G^+ (z) &= - e^{-2\rho-i\sigma} (p)^4 
+ {\textstyle{i\over 2}}e^{-\rho} p_a p_b
\partial x^{ab}\cr 
&\hskip 10pt
+ e^{i\sigma} (-\half\partial x^m\partial x_m - p_a \partial\theta^a 
-\half\partial (\rho + i\sigma) \partial (\rho + i\sigma) 
+\half\partial^2  (\rho + i\sigma) ) \,+ G^+_C\cr
G^- (z) &= e^{-i\sigma} + G_C^- \cr
J (z) &=  \partial (\rho + i\sigma) + J_C\cr
J^+ (z) &=  e^{\rho + i\sigma} J_C^+ = - e^{\rho + i\sigma +i H_C}\cr
J^- (z) &=  - e^{-\rho - i\sigma} J_C^- = - e^{-\rho - i\sigma - i H_C}\cr
\tilde G^+(z) &= e^{iH_C + \rho} + e^{\rho + i\sigma} \tilde G^+_C\cr
\tilde G^-(z) &= e^{-iH_C} [\,-e^{-3\rho-2i\sigma} (p)^4 
- {\textstyle{i\over 2}}
e^{-2\rho-i\sigma} p_a p_b \partial x^{ab}\cr
&\hskip 10pt 
+ e^{-\rho} (-\half\partial x^m\partial x_m 
- p_a \partial\theta^a -\half\partial (\rho + i\sigma) 
\partial (\rho + i\sigma)
+\half\partial^2  (\rho + i\sigma) ) \,]\cr
&\hskip 10pt - e^{-\rho-i\sigma} \tilde G_C^-\,.\cr}}
These currents are given in terms of the left-moving bosons
$\partial x^m, \rho, \sigma$, and the left-moving fermionic worldsheet
fields $p_a,\t^a$,
where $0\le m\le 5; 1\le a\le 4$. 
The conformal weights of $p_a,\t^a$ are $1$ and $0$, respectively.
The BVW variables no longer exhibit the matter times
ghost sector structure familiar from the conventional
Ramond-Neveu-Schwarz formalism, with cancelling contributions
of $\pm 15$ to the central charge. In BVW, the residual ghost
fields are $\rho,\sigma$. (In the ten-dimensional version,
these are promoted to
complex worldsheet boson fields $\lambda^\alpha$ with
a spacetime Majorana spinor index $\alpha$, which are
parameterized by eleven complex fields \refs{\btenflat-\blicoh}.)
We define $p^4\equiv {\scriptstyle{1\over 24}}
\epsilon^{abcd}  p_a p_b p_c p_d = p_1p_2 p_3 p_4;$
and $\partial x^{ab} = \partial x^m\,\sigma_m^{ab}$ where
the sigma matrices in flat space satisfy 
$\sigma_m^{ab} \,\sigma_{nac} \,+ \sigma_n^{ab} \,\sigma_{mac}
= \eta_{mn}\,\delta^b_c $. Here lowered indices mean 
$\sigma_{mab}\equiv \half\epsilon_{abcd} \sigma_m^{cd}.$ 
Note that $e^\rho$ and $e^{i\sigma}$ are worldsheet fermions.
Also $e^{\rho + i\sigma}\equiv e^\rho e^{i\sigma} = - e^{i\sigma} e^\rho$.
Here $J_C\equiv i\partial H_C, J_C^+ \equiv - e^{i H_C},
J_C^-\equiv e^{-i H_C}$.
Both $\tilde T, G^\pm, J, J^\pm, \tilde G^\pm$
and the generators describing the $K3$ compactification 
$\tilde T_C, G_C^\pm, J_C, J_C^\pm, \tilde G_C^\pm$ satisfy the
twisted $N=4$, $c=6$,  superconformal algebra, {\it i.e.} both
$\tilde T$ and $\tilde T_C$ have zero central charge.
However, $c$ still appears in the twisted $N=4$ and $N=2$ algebras
in the products involving the supercurrents and the SU(2) currents; 
and the N=2 generators in \svgen\ $\tilde T, G^\pm, J$
decompose into a $c=0$ 
six-dimensional part and a $c=6$ compactification-dependent piece. 

The other non-vanishing OPE's are
$x^m(z,\bar z) x^n(\zeta,\bar\zeta) = -\eta^{mn}\ln |z-\zeta|$; 
for the left-moving worldsheet fermion fields 
$\,p_a (z) \theta^b (\zeta) = (z-\zeta)^{-1}\delta_a^b$; and for the
left-moving worldsheet bosons 
$\rho (z) \rho(\zeta) = - \ln (z-\zeta)\,;\,\,$
$\sigma (z) \sigma (\zeta) = -\ln (z-\zeta)\,.$ 
Right-movers are denoted by barred notation and have similar OPE's.

\subsec{$N=4$ Superconformal Algebra for $AdS_3\times S^3$}

We also recall \refs{\bvw}\ the expression for the twisted holomorphic
$N=4$ generators when ${\bf R^6}$ is replaced by $AdS_3\times S^3$
in the presence of Ramond flux:
\eqn\svaads{
\eqalign{\tilde T (z) &
= -{\textstyle{1\over 8}} \epsilon^{abcd} \, K_{ab} K_{cd} 
- F^a E^a 
-\half\partial\rho\partial\rho - \half\partial\sigma\partial\sigma
+ \thalf\partial^2 (\rho + i\sigma) + \tilde T_C\cr
G^+ (z) &= - {\textstyle{1\over 6}}e^{-2\rho-i\sigma} \epsilon^{abcd}
U_{ab} U_{cd} 
+ i e^{-\rho} K^{ab} U_{ab}\cr
&\hskip 10pt
+ e^{i\sigma} (-{\textstyle{1\over 8}} \epsilon^{abcd} \, K_{ab} K_{cd}
- F^a E^a 
-\half\partial (\rho + i\sigma) \partial (\rho + i\sigma)
+\half\partial^2  (\rho + i\sigma) ) \,+ G^+_C\cr
G^- (z) &= e^{-i\sigma} + G_C^- \cr
J (z) &=  \partial (\rho + i\sigma) + J_C\cr
\tilde G^+(z) &= e^{iH_C + \rho} + e^{\rho + i\sigma} \tilde G^+_C\,,\cr}}
where  \eqn\uab
{U_{ab} = ( 1 - {\textstyle{1\over 4}} e^{\varphi + \bar\varphi} )^{-2}
\, [ \half F_a F_b -{\textstyle{1\over 8}} e^{2\bar\varphi} E_a E_b
-{\textstyle{i\over 4}} e^{\bar\varphi} ( E_a E_b + F_a F_b )\,]\,,}
and the remaining generators $J^\pm, \tilde G^-$ can be constructed from
$\tilde T, G^\pm, J$ \refs{\bvw}.
Here $ e^\varphi \equiv e^{-\rho - i \sigma}\,,$
and $ F_a, E_a, K_{ab} $ are the fermionic and bosonic z-components
of the right invariant $PSU(2|2)$ currents. There is a corresponding
anti-holomorphic $N=4$ algebra in terms of barred worldsheet fields.
Although the $N=4$ generators have definite holomorphicity, the
worldsheet fields $F_a, E_a, K_{ab}, \bar F_a, \bar E_a, \bar K_{ab}$
do not. They are each functions of $z$ and $\bar z$ and have
non-free operator products, which are not known in closed form. They reduce  
to the free conformal fields
$p_a, \partial\theta_a, \partial x^{ab}, \bar p_a, 
\bar\partial\bar\theta_a, \bar \partial x^{ab}$ only in the flat limit.

\newsec{Three-graviton tree amplitude in flat space}

We first compute the six-dimensional
three-graviton tree level amplitude in (6d) flat space,
for Type IIB superstrings on ${\bf R^6}\times K3$
in the BVW formalism. It is contained in the closed string
three-point function
\eqn\sam{<\,V (z_1,\bar z_1)\,( G^+_0 \bar G^+_0  V (z_2, \bar z_2))
(\tilde G^+_0 \bar{\tilde G}^+_0 V (z_3, \bar z_3)) \,>\,}
where the $N=4$ supercurrents are found in \svgen\ , and  
the vertex operators are given by
\eqn\gravvo{V (z,\bar z) = e^{i\sigma(z) + \rho(z)}\,
e^{i\bar\sigma(\bar z) + \bar\rho(\bar z)}\,
\theta^a(z) \theta^b(z) \,\bar\theta^{\bar a} (\bar z)
\,\bar\theta^{\bar b} (\bar z) \,\sigma_{ab}^m \,\sigma_{\bar a\bar b}^n
\,\phi_{mn}(X(z,\bar z))\,,} 
when the field $$\phi_{mn} = g_{mn} + b_{mn} + \bar g_{mn}\,\phi\,$$
satisfies the constraint equations \refs{\bvw,\dw}
$\partial^m\phi_{mn} = 0$, and $\Box \phi_{mn} = 0\,.$
These constraints imply the gauge conditions $\p^m b_{mn} = 0$ 
for the two-form, and
$\p^m g_{mn} = -\p_n \phi$ for the traceless graviton $g_{mn}$
and dilaton $\phi\,.$
The constraints follow from the physical state conditions
which in this formalism are implemented by the $N=4$ generators,
as shown in \bvw. ( Since the $N=4$ algebra is twisted, {\it i.e.}
topological, the nilpotent generators
$G^+,\tilde G^+, \bar G^+,\tilde{\bar G}^+$
are dimension one, and their cohomology essentially
determines the physical states.)
There is a residual gauge symmetry
\eqn\residgt{g_{mn}\rightarrow g_{mn} + \p_m\xi_n + \p_n\xi_m \,,
\qquad \phi\rightarrow \phi\,,\qquad b_{mn}\rightarrow
b_{mn}\,}
with $\Box \xi_n = 0\,,\, \p\cdot\xi = 0\,.$
To evaluate \sam, we first extract the simple poles
\eqn\gonv{\eqalign{G_0^+ \bar G_0^+ \, V (z,\bar z) = e^{i\sigma }\,
e^{i\bar\sigma}\,(-4) \,&
\,[ \phi_{mn}(X) \,\partial X^m \bar\partial X^n \cr
&-p_a \theta^b \,\sigma_{cb}^m\,\sigma^{p c a}
\bar\partial X^n \,\partial_p\,\phi_{mn}(X)\cr
&-\bar p_{\bar a} \bar\theta^{\bar b} \,\sigma_{\bar c\bar b}^n\,
\sigma^{p \bar c\bar a}
\partial X^m \,\partial_p \phi_{mn}(X)\cr
&+ p_a \theta^b \,\bar p_{\bar a} \bar\theta^{\bar b} \,
\sigma_{cb}^m\,\sigma^{p c a}\,
\sigma_{\bar c\bar b}^n\, \sigma^{q \bar c\bar a}\,
\partial_p\partial_q \,\phi_{mn}(X)\,]\,,\cr}}
\eqn\gtonv{\tilde G_{0}^+ \bar {\tilde G}_{0}^+
\, V (z,\bar z) = e^{i H_C + 2\rho + i\sigma }\,
e^{i\bar H_C + 2\bar\rho + i\bar\sigma}\,
\theta^a \theta^b \,\bar\theta^{\bar a}
\,\bar\theta^{\bar b} \,\sigma_{ab}^m \,\sigma_{\bar a\bar b}^n
\,\phi_{mn}(X)\,.}
Then, using the OPE's for the ghost fields and $H_C$, we partially
compute \sam\ by exhibiting their contribution as \refs{\bpz-\cf} 
$$\eqalignno{&<\,V_1(z_1,\bar z_1)\,( G^+_0 \bar G^+_0  V_2(z_2, \bar z_2))
(\tilde G^+_0 \bar{\tilde G}^+_0 V_3(z_3, \bar z_3)) \,>\cr
&= (z_1-z_2) (z_2-z_3) (z_1-z_3)^{-1}
(\bar z_1 - \bar z_2) (\bar z_2 - \bar z_3) (\bar z_1 - \bar z_3)^{-1}\cr
& \hskip8pt\cdot 
4 <\, e^{iH_C(z_3)} e^{\rho (z_3) + 2\rho(z_3)} e^{3i\sigma (z_3)}\,
e^{i\bar H_C(\bar z_3)} e^{\bar\rho (\bar z_3) + 2\bar\rho(\bar z_3)}
e^{3i\bar\sigma (\bar z_3)}\cr
&\hskip8pt\cdot \, \theta^a(z_1) \theta^b(z_1) \bar\theta^{\bar a}(\bar z_1)
\bar\theta^{\bar b}(\bar z_1) \,\sigma_{ab}^m \,\sigma_{\bar a\bar b}^n
\,\phi_{mn}(X(z_1,\bar z_1))\,\cr
&\hskip20pt\cdot \,[\,\phi_{jk}(X(z_2,\bar z_2))\,
\partial X^j(z_2) \bar\partial X^k(\bar z_2)\cr
&\hskip25pt
- p_e(z_2) \theta^f(z_2) \,\sigma_{uf}^j\,\sigma^{p u e}
\bar\partial X^k (\bar z_2)\,\partial_p\,\phi_{jk}(X(z_2,\bar z_2))\cr
&\hskip25pt
-\bar p_{\bar e}(\bar z_2)
\bar\theta^{\bar f}(\bar z_2) \,\sigma_{\bar u\bar f}^k\,
\sigma^{p \bar u\bar e}
\partial X^j (z_2)\,\partial_p \phi_{jk}(X(z_2,\bar z_2))\cr
&\hskip25pt
+ p_e(z_2)\theta^f (z_2)\,
\bar p_{\bar e}(\bar z_2) \bar\theta^{\bar f}(\bar z_2) \,
\sigma_{uf}^j\,\sigma^{p u e}\,
\sigma_{\bar u\bar f}^k\, \sigma^{q \bar u\bar e}\,
\partial_p\partial_q \,\phi_{jk}(X(z_2,\bar z_2))\,]\cr
&\hskip20pt\cdot\,\theta^c(z_3) \theta^d(z_3) \bar\theta^{\bar c}(\bar z_3)
\bar\theta^{\bar d}(\bar z_3) \,\sigma_{cd}^g \,\sigma_{\bar c\bar d}^h
\,\phi_{gh}(X(z_3,\bar z_3))\,>\cr}$$
Computing the remaining $z_2, z_3$ operators products, and
using the $SL(2,C)$ invariance of the amplitude to take the
three points to constants
$z_1\rightarrow\infty$, $\bar z_1\rightarrow\infty$,
$z_2\rightarrow 1, \bar z_2\rightarrow 1$, 
$z_3\rightarrow 0, \bar z_3\rightarrow 0$,
we have
$$\eqalignno{&<\,V_1(z_1,\bar z_1)\,( G^+_0 \bar G^+_0  V_2(z_2, \bar z_2))
(\tilde G^+_0 \bar{\tilde G}^+_0 V_3(z_3, \bar z_3)) \,>\cr
&= (z_2-z_3) (\bar z_2 - \bar z_3) (z_2-z_3)^{-1} (\bar z_2 - \bar z_3)^{-1}
\cdot 4\cr
& \hskip8pt\cdot\,<\, e^{iH_C(0) +3\rho(0) + 3i\sigma(0)}
e^{i\bar H_C(0) + 3\bar\rho(0) + 3i\bar\sigma(0)}
\theta^a_0\theta^b_0\theta^c_0\theta^d_0\,
\bar\theta^{\bar a}_0
\bar\theta^{\bar b}_0\bar\theta^{\bar c}_0\bar\theta^{\bar d}_0\,>\cr
& \hskip8pt\cdot\,[ \sigma_{ab}^m \,\sigma_{cd}^g \,
\sigma_{\bar a\bar b}^n\, \sigma_{\bar c\bar d}^h
\, \,<\phi_{mn}(X(\infty))\,
\phi_{jk}(X(1)) \,\partial^j\partial^k\phi_{gh}(X(0))>\cr
&\hskip17pt +2 \sigma_{ab}^m \, (\sigma^j\sigma^p\sigma^g)_{cd}\,
\sigma_{\bar a\bar b}^n\, \sigma_{\bar c\bar d}^h\,
\,<\phi_{mn}(X(\infty))\,
\partial_p \phi_{jk}(X(1))
\, \partial^k\phi_{gh}(X(0))>\cr
&\hskip17pt +2 \sigma_{ab}^m \, \sigma^g_{cd}\,
\sigma_{\bar a\bar b}^n\, (\sigma^k\sigma^p\sigma^h)_{\bar c\bar d}\,
\,<\phi_{mn}(X(\infty))\,
\partial_p \phi_{jk}(X(1))
\, \partial^j\phi_{gh}(X(0))>\cr
&\hskip17pt +4 \sigma_{ab}^m \, (\sigma^j\sigma^p\sigma^g)_{cd}\,
\sigma_{\bar a\bar b}^n\, (\sigma^k\sigma^q\sigma^h)_{\bar c\bar d}\,
\,<\phi_{mn}(X(\infty))\,
\partial_p\partial_q \phi_{jk}(X(1))
\,\phi_{gh}(X(0))>]\cr
&=4\,[\,\,\bar g^{mg}\bar g^{nh}
\, \,<\phi_{mn}(x_0)\,
\phi_{jk}(x_0) \,\partial^j\partial^k\phi_{gh}(x_0)>\cr
&\hskip17pt - \bar g^{nh} \,
(\sigma^m\sigma^j\sigma^p\sigma^g)^d_{\hskip2pt d}\,\,<\phi_{mn}(x_0)\,
\partial_p \phi_{jk}(x_0)
\, \partial^k\phi_{gh}(x_0)>\cr
&\hskip17pt - \bar g^{mg}\,
(\sigma^n\sigma^k\sigma^p\sigma^h)^{\bar d}_{\hskip2pt\bar d}\,
\,<\phi_{mn}(x_0)\,
\partial_p \phi_{jk}(x_0)
\, \partial^j\phi_{gh}(x_0)>\cr
&\hskip17pt + (\sigma^m\sigma^j\sigma^p\sigma^g)^d_{\hskip2pt d}\,\,
(\sigma^n\sigma^k\sigma^q\sigma^h)^{\bar d}_{\hskip2pt\bar d}\,
\,<\phi_{mn}(x_0)\,
\partial_p\partial_q \phi_{jk}(x_0)
\,\phi_{gh}(x_0)>]\cr
}$$
where the second equality follows from the vacuum expectation value
of the ghost fields, $H_C$ and eight fermion zero modes
$<\, e^{iH_C(0) +3\rho(0) + 3i\sigma(0)}
e^{i\bar H_C(0) + 3\bar\rho(0) + 3i\bar\sigma(0)}
\theta^a_0\theta^b_0\theta^c_0\theta^d_0\,
\bar\theta^{\bar a}_0
\bar\theta^{\bar b}_0\bar\theta^{\bar c}_0\bar\theta^{\bar d}_0\,>
 = {\textstyle{1\over 16}}
\epsilon^{abcd}\,\epsilon^{\bar a\bar b\bar c\bar d}$,
and various sigma matrix identities \refs{\bvw-\dw}.
Further using $(\sigma^m\sigma^n\sigma^p\sigma^q)^d_{\hskip2pt d}
= \bar g^{mn}\bar g^{pq} + \bar g^{mq}\bar g^{np}
- \bar g^{mp}\bar g^{nq}$ where in flat space $\bar g_{mn} = \eta_{mn}$,
we have
$$\eqalignno{&<\,V_1(z_1,\bar z_1)\,( G^+_0 \bar G^+_0  V_2(z_2, \bar z_2))
(\tilde G^+_0 \bar{\tilde G}^+_0 V_3(z_3, \bar z_3)) \,>\cr
&=4\,[\,\,\bar g^{mg}\bar g^{nh}
\, \,<\phi_{mn}(x_0)\,
\phi_{jk}(x_0) \,\partial^j\partial^k\phi_{gh}(x_0)>\cr
&\hskip17pt - \bar g^{nh} (\bar g^{mj}\bar g^{pg} + \bar g^{mg}\bar g^{jp}
- \bar g^{mp}\bar g^{jg} )\,
\,<\phi_{mn}(x_0)\,
\partial_p \phi_{jk}(x_0)
\, \partial^k\phi_{gh}(x_0)>\cr
&\hskip17pt - \bar g^{mg}\,
(\bar g^{nk}\bar g^{ph} + \bar g^{nh}\bar g^{kp} - \bar g^{np}\bar g^{kh} )\,
\,<\phi_{mn}(x_0)\,
\partial_p \phi_{jk}(x_0)
\, \partial^j\phi_{gh}(x_0)>\cr
&\hskip17pt + (\bar g^{mj}\bar g^{pg} + \bar g^{mg}\bar g^{jp}
- \bar g^{mp}\bar g^{jg} )\,
(\bar g^{nk}\bar g^{qh} + \bar g^{nh}\bar g^{kq} - \bar g^{nq}\bar g^{kh} )\cr
&\hskip25pt \cdot <\phi_{mn}(x_0)\,
\partial_p\partial_q \phi_{jk}(x_0)
\,\phi_{gh}(x_0)>]\,.\cr}$$
Using the gauge condition $\partial^m\phi_{mn} = 0$ again, then finally 
\eqn\thrgrav{\eqalign
{&<\,V_1(z_1,\bar z_1)\,( G^+_0 \bar G^+_0  V_2(z_2, \bar z_2))
(\tilde G^+_0 \bar{\tilde G}^+_0 V_3(z_3, \bar z_3)) \,>\cr
&=4\,[\,\,\bar g^{mg}\bar g^{nh}
\, \,<\phi_{mn}(x_0)\,
\phi_{jk}(x_0) \,\partial^j\partial^k\phi_{gh}(x_0)>\cr
&\hskip17pt - \bar g^{nh} (\bar g^{mj}\bar g^{pg} 
- \bar g^{mp}\bar g^{jg} )\,
\,<\phi_{mn}(x_0)\,
\partial_p \phi_{jk}(x_0)
\, \partial^k\phi_{gh}(x_0)>\cr
&\hskip17pt - \bar g^{mg}\,
(\bar g^{nk}\bar g^{ph} - \bar g^{np}\bar g^{kh} )\,
\,<\phi_{mn}(x_0)\,
\partial_p \phi_{jk}(x_0)
\, \partial^j\phi_{gh}(x_0)>\cr
&\hskip17pt + (\bar g^{mj}\bar g^{pg} 
- \bar g^{mp}\bar g^{jg} )\,
(\bar g^{nk}\bar g^{qh} - \bar g^{nq}\bar g^{kh} ))\cr
&\hskip25pt \cdot <\phi_{mn}(x_0)\,
\partial_p\partial_q \phi_{jk}(x_0)
\,\phi_{gh}(x_0)>]\cr
&= 12\,[\, <\phi^{mn}(x_0)\,\phi^{jk}(x_0)\,
\partial_m\partial_n \phi_{jk}(x_0)>
+ 2\, <\phi^{mn}(x_0)\,\partial_m\phi^{jk}(x_0)\,
\partial_j \phi_{nk}(x_0)> \,]\,.\cr}}
To make contact with the supergravity field theory expression,
we can evaluate this dual model amplitude in either
momentum space or in position space. For this flat space case,
momentum space is often used, since momentum is conserved,
{\it i.e.} $k_1 + k_2 + k_3 = 0$. For AdS, momentum is not conserved,
so we will just work in the position space representation
in both cases. 

The Einstein-Hilbert action is 
$I = \int d^dx {\sqrt{|g|}} \{-{R\over{2 K^2}}\}\,.$
Expanding to third order in $K$ using 
$g_{\mu\nu} = \eta_{\mu\nu} + 2 K h_{\mu\nu}$, we find 
the three-point interaction $I_3$.
In harmonic gauge, {\it i.e.} when
$\p^\mu h_{\mu\nu} - \half \p_\nu h^\rho_\rho = 0$, and
on shell $\Box h_{\mu\nu} = 0$, it is given by 
\eqn\flat{
I_3 = - K \int d^dx [ h^{\mu\nu} h^{\rho\sigma} \partial_\mu\partial_\nu
h_{\rho\sigma} + 2 h^{\mu\nu} \partial_\mu h^{\rho\sigma} \partial_\rho
h_{\nu\sigma} ]\,.}
The gauge transformations
\eqn\gtran{h_{\mu\nu} \rightarrow h_{\mu\nu} + \p_\mu\xi_\nu 
+ \p_\nu\xi_\mu}
leave invariant the harmonic gauge condition and
$I_3$, given in \flat , when $\Box\xi_\mu = 0\,.$
Using this gauge symmetry, we could further choose
$h^\rho_\rho = 0$, $\p^\mu h_{\mu\nu} = 0$. 
Then $I_3$ represents the three-graviton amplitude, and  
is invariant under residual gauge transformations
that have $\p\cdot\xi = 0$.

To extract the string theory three-graviton amplitude 
from \thrgrav, we set $b_{mn}$ to zero, and
use the field identifications\dw\ that relate
the string fields $g_{mn}, \phi$ to the supergravity
field $h_{mn}$ via $\phi\equiv -{\textstyle{1\over 3}}h^\rho_\rho$
and $g_{mn}\equiv h_{mn} - {\textstyle{1\over 6}} \bar g_{mn}
h^\rho_\rho$, where $h_{mn}$ is in harmonic gauge. 
Then $\phi_{mn} = h_{mn} -\half\bar g_{mn}
h^\rho_\rho\,,$ and from \thrgrav\ the on shell string tree amplitude is  
\eqn\thrgravft{\eqalign
{&{\textstyle{-{K\over {12}}}}
<\,V_1(z_1,\bar z_1)\,( G^+_0 \bar G^+_0  V_2(z_2, \bar z_2))
(\tilde G^+_0 \bar{\tilde G}^+_0 V_3(z_3, \bar z_3)) \,>\cr
&= -K \int d^d x \, [ \, \phi^{mn}(x)\,\phi^{jk}(x)\,
\partial_m\partial_n \phi_{jk}(x)
+ 2\, \phi^{mn}(x)\,\partial_m\phi^{jk}(x)\,
\partial_j \phi_{nk}(x) \,]\cr
&=  - K \int d^d x \, [ h^{mn} h^{jk}\, \p_m \p_n h_{jk}\,
+ 2\, h^{mn} \p_m h^{jk} \p_j h_{nk}\,]
\,+ K  \int d^d x \, h^{mn} \,\p_m h^k_k \,\,\p_n h^p_p  \cr
&= I_3  + I'_3}}
where $I'_3$ is the one graviton - two dilaton amplitude,
$I_3$ is the three graviton interaction in harmonic gauge,
and $d=6$. 
We note that $I_3$ and $I'_3$ separately are invariant
under the gauge transformation \gtran\ with
$\Box \xi_n = 0$ and $\p\cdot\xi = 0$, which corresponds to
the gauge symmetry of the string field
$\phi_{mn}\rightarrow \phi_{mn} + \p_m\xi_n + \p_n\xi_m\,.$
Furthermore, $I_3$ is also invariant under gauge transformations
for which $\p\cdot\xi\ne 0$, and these can be used to
eliminate the trace of $h_{mn}$ in $I_3$.
In the string gauge, the trace of $\phi_{mn}$ is related to 
the dilaton $\phi^m_m =  6\phi $, so even when $b_{mn} = 0$, \thrgrav\ 
contains both the three graviton amplitude and the one graviton - two
dilaton interaction. 
{}From \thrgravft\ we see that
we could have extracted $I_3$ from \thrgrav\ merely by
setting both $b_{mn}=0$ and $\phi = 0$, since then $\phi_{mn} = g_{mn}$
and $\p^m g_{mn} = 0\,.$

\newsec{Three-graviton tree amplitude in $AdS_3\times S^3$}

In this section, we compute the six-dimensional 
three-graviton amplitude in the Type IIB superstring on 
$AdS_3\times {\bf S^3}\times K3$ with background Ramond flux. Consider
\eqn\thrgravads{<\,V_1(z_1,\bar z_1)\,( G^+_0 \bar G^+_0  V_2(z_2, \bar z_2))
(\tilde G^+_0 \bar{\tilde G}^+_0 V_3(z_3, \bar z_3)) \,>}
where the $N=4$ supercurrents are reviewed in \svaads,\uab.
They are expressed in terms of the left action generators
defined in \dw\ as
\eqn\ssg{\eqalign{F_a &= {d\over d\t^a}\,,\qquad
K_{ab} = -\t_a {d\over d\t^b} + \t_b {d\over d\t^a} + t_{Lab}\cr
E_a &= {\textstyle{1\over 2}}\epsilon_{abcd}\,\t^b\,
( t_{L}^{cd} - \t^c {d\over d\t_d}\,) + h_{a\bar b} 
{d\over d\bar\t_{\bar b}}\,,\cr}}
and corresponding right action generators $\bar F_a, \bar E_a,
\bar K_{ab}$.
The vertex operators are 
$$V (z,\bar z) = e^{i\sigma(z) + \rho(z)}\,
e^{i\bar\sigma(\bar z) + \bar\rho(\bar z)}\,
\theta^a(z,\bar z) \theta^b(z,\bar z) \,\bar\theta^{\bar a} (z, \bar z)
\,\bar\theta^{\bar b} (z, \bar z) \,\sigma_{ab}^m \,\sigma_{\bar a\bar b}^n
\,\phi_{mn}(X(z,\bar z))\,.$$
On $AdS$, in addition to the equation of motion, 
the string field $\phi_{mn} = g_{mn} + b_{mn} + \bar g_{mn}\,\phi\,$
satisfies constraints given by
$t^{ab}_L\, h^g_{\,\,\bar a}\,  
h^h_{\,\,\bar b}\,\sigma^m_{ab}\,\sigma^n_{gh}\,\phi_{mn} =\,0\,,\,
t^{\bar a\bar b}_R\, h^{\,\,\bar g}_a\,  h^{\,\,\bar h}_{b}\,
\sigma^m_{\bar g\bar h}\,\sigma^n_{\bar a \bar b}\,\phi_{mn}
=0$, which were derived in \dw\ where
$t^{ab}_L, t^{\bar a\bar b}_R$  describe invariant derivatives on the $SO(4)$ 
group manifold. These can be related to covariant derivatives
${\cal T}_L^{cd} \equiv -\sigma^{p\,cd} \, D_p\,,\,
{\cal T}_R^{\bar c\bar d} \equiv \sigma^{p\, \bar c\bar d} \, D_p\,\,,$
where for example, acting on a function,
${\cal T}_L=t_L$ and ${\cal T}_R=t_R$. But 
when acting acting on fields that carry vector or spinor indices,
they differ so that for example on spinor indices
$t_L^{ab} V_e=
{\cal T}_L^{ab} \, V_e + \half \delta^a_e\, \delta^{bc} V_c
- \half \delta^b_e\, \delta^{ac} V_c\,.$
In terms of covariant derivatives,
the constraints are
$D^m g_{mn} = D^m g_{nm} = - D_n\phi \,+ \,\bar H_{nrs} b^{rs}\,,$
and $D^m b_{mn} = 0 = D^m b_{nm}\,.$
These are the $AdS\times S $ analog of the flat space
constraints $\partial^m\phi_{mn} = 0\,.$
On $AdS_3\times S^3$, 
\eqn\gzero{\eqalign{G_0^+ \bar G_0^+& \, V (z,\bar z)\cr
 = e^{i\sigma }&\,
e^{i\bar\sigma}\,(-4)\cr
\cdot& [ \,{\textstyle{1\over 4}} K^{ab}(z,\bar z)\,
\bar K^{\bar a\bar b}(z,\bar z)\,
\sigma^m_{ab}\,\sigma^n_{\bar a\bar b}\,\phi_{mn}(X)\cr
&-{\textstyle{1\over 2}} F_a (z,\bar z)\,
\theta^b(z,\bar z)\, \,
\bar K^{\bar a\bar b}(z,\bar z)\,
(t_L^{ac} - \delta^{ac})\,
\sigma^m_{cb}\,\sigma^n_{\bar a\bar b}\,\phi_{mn}(X)\cr
&-{\textstyle{1\over 2}} \bar F_{\bar a} (z,\bar z)\,
\bar\theta^{\bar b}(z,\bar z)\, \,
K^{ab}(z,\bar z)\,
(t_R^{\bar a\bar c} - \delta^{\bar a\bar c})\,
\sigma^m_{ab}\,\sigma^n_{\bar c\bar b}\,\phi_{mn}(X)\cr
&+ F_a (z,\bar z)\,\theta^b(z,\bar z)\, \,
\bar F_{\bar a} (z,\bar z)\,\bar\theta^{\bar b}(z ,\bar z)\, \,
(t_R^{\bar a\bar c} - \delta^{\bar a\bar c})\,
(t_L^{ac} - \delta^{ac})\,
\sigma^m_{cb}\,\sigma^n_{\bar c\bar b}\,\phi_{mn}(X)\,]\cr}}
\eqn\gtildezero{\hskip-30pt\tilde G_{0}^+ \bar {\tilde G}_{0}^+
\, V (z,\bar z) = e^{i H_C + 2\rho + i\sigma }\,
e^{i\bar H_C + 2\bar\rho + i\bar\sigma}\,
\theta^a \theta^b \,\bar\theta^{\bar a}
\,\bar\theta^{\bar b} \,\sigma_{ab}^m \,\sigma_{\bar a\bar b}^n
\,\phi_{mn}(X)\,.}
In deriving \gzero, we keep only the contribution
$i e^{-\rho} K^{ab} U_{ab}$ to $G^+$ with $U_{ab}\sim \half
F_a F_b$ since other terms in $G^+$ do not survive the ghost measure 
in the vacuum expectation value.
We have also assumed that the OPE of $F_a (z,\bar z)$ with 
$\theta^e(\zeta,\bar\zeta)$ can be replaced with
$F_a (z,\bar z) \theta^e (\zeta,\bar\zeta) \sim (z-\zeta)^{-1}
\delta_a^e\,,$ in accordance with \ssg.
This is motivated by the observation that
evaluating the OPE's in this manner leads to the
constraint equations found in \dw, where 
those equations were derived solely by requiring supersymmetric invariance
(and not from the action of the $N=4$ generators).
\eqn\begads
{\eqalign{&<\,V_1(z_1,\bar z_1)\,( G^+_0 \bar G^+_0  V_2(z_2, \bar z_2))
(\tilde G^+_0 \bar{\tilde G}^+_0 V_3(z_3, \bar z_3)) \,>\cr
&= (z_1-z_2) (z_2-z_3) (z_1-z_3)^{-1}
(\bar z_1 - \bar z_2) (\bar z_2 - \bar z_3) (\bar z_1 - \bar z_3)^{-1}\cr
& \hskip8pt\cdot 4 <\, e^{iH_C(z_3)} e^{\rho (z_3) + 2\rho(z_3)} 
e^{3i\sigma (z_3)}\,
e^{i\bar H_C(\bar z_3)} e^{\bar\rho (\bar z_3) + 2\bar\rho(\bar z_3)}
e^{3i\bar\sigma (\bar z_3)}\cr
&\hskip8pt\cdot \, \theta^a(z_1) \theta^b(z_1) \bar\theta^{\bar a}(\bar z_1)
\bar\theta^{\bar b}(\bar z_1) \,\sigma_{ab}^m \,\sigma_{\bar a\bar b}^n
\,\phi_{mn}(X(z_1,\bar z_1))\,\cr
&\hskip20pt\cdot \,[\,{\textstyle{1\over 4}}\,\phi_{jk}(X(z_2,\bar z_2))\,
\sigma^j_{ef}\,\sigma^k_{\bar e\bar f}\,
K^{ef}(z_2,\bar z_2)\,
\bar K^{\bar e\bar f}(z_2,\bar z_2)\cr
&\hskip25pt
-{\textstyle{1\over 2}} F_e (z_2,\bar z_2)\,
\theta^f(z_2,\bar z_2)\, \,
\bar K^{\bar e\bar f}(z_2,\bar z_2)\,
(t_L^{e\ell} - \delta^{e\ell})\,
\sigma^j_{\ell f}\,\sigma^k_{\bar e\bar f}\,\phi_{jk}(X)\cr
&\hskip25pt
-{\textstyle{1\over 2}} \bar F_{\bar e} (z_2,\bar z_2)\,
\bar\theta^{\bar f}(z_2,\bar z_2)\, \,
K^{ef}(z_2,\bar z_2)\,
(t_R^{\bar e\bar\ell} - \delta^{\bar e\bar\ell})\,
\sigma^j_{ef}\,\sigma^k_{\bar\ell\bar f}\,\phi_{jk}(X)\cr
&\hskip25pt
+F_e (z_2,\bar z_2)\,\theta^f(z_2,\bar z_2)\, \,
\bar F_{\bar e} (z_2,\bar z_2)\,\bar\theta^{\bar f}(z_2,\bar z_2)\, \,
(t_R^{\bar e\bar \ell} - \delta^{\bar e\bar \ell})\,
(t_L^{e\ell} - \delta^{e\ell})\,
\sigma^j_{\ell f}\,\sigma^k_{\bar\ell\bar f}\,\phi_{jk}(X)\,]\cr
&\hskip20pt\cdot\,\theta^c(z_3) \theta^d(z_3) \bar\theta^{\bar c}(\bar z_3)
\bar\theta^{\bar d}(\bar z_3) \,\sigma_{cd}^g \,\sigma_{\bar c\bar d}^h
\,\phi_{gh}(X(z_3,\bar z_3))\,>\,.\cr}}
Evaluating at $z_1\rightarrow\infty$, $\bar z_1\rightarrow\infty$,
and restricting $\phi_{mn}, \phi_{jk}, \phi_{gh}$ to be symmetric, we have
\eqn\medads{\eqalign
{&<\,V_1(z_1,\bar z_1)\,( G^+_0 \bar G^+_0  V_2(z_2, \bar z_2))
(\tilde G^+_0 \bar{\tilde G}^+_0 V_3(z_3, \bar z_3)) \,>\cr
&= (z_2-z_3) (\bar z_2 - \bar z_3) (z_2-z_3)^{-1} (\bar z_2 - \bar z_3)^{-1}
\cdot 4\cr
& \hskip8pt\cdot\,<\, e^{iH_C(0) +3\rho(0) + 3i\sigma(0)}
e^{i\bar H_C(0) + 3\bar\rho(0) + 3i\bar\sigma(0)}
\theta^a_0\theta^b_0\theta^c_0\theta^d_0\,
\bar\theta^{\bar a}_0
\bar\theta^{\bar b}_0\bar\theta^{\bar c}_0\bar\theta^{\bar d}_0\,>
\,\leftarrow {\textstyle{1\over 16}}
\epsilon^{abcd}\,\epsilon^{\bar a\bar b\bar c\bar d}\cr
& \hskip8pt\cdot\,[ \,{\textstyle {1\over 4}} \sigma_{ab}^m \,
\sigma_{\bar a\bar b}^n\,
\sigma^j_{ef}\,\sigma^k_{\bar e\bar f}
\, \,<\phi_{mn}(X(\infty))\, \phi_{jk}(X(1)) \,
[\,t_L^{ef}\, t_R^{\bar e\bar f}\,\,\sigma_{cd}^g \sigma_{\bar c\bar d}^h
 \,\phi_{gh}(X(0))\,]\,>\cr
&\hskip17pt + \sigma_{ab}^m \,\sigma_{\bar a\bar b}^n\,\, \,
<\phi_{mn}(X(\infty))\,
[\, t_L^{e\ell} \sigma^j_{\ell c} \sigma^k_{\bar e\bar f} \, 
\phi_{jk}(X(1)) ]\,[\, t_R^{\bar e\bar f} \sigma^g_{ed} 
\sigma^h_{\bar c \bar d}
\,\phi_{gh}(X(0))\,]\,>\cr
&\hskip17pt + \sigma_{ab}^m \,\sigma_{\bar a\bar b}^n\,\, \,
<\phi_{mn}(X(\infty))\,
[\, t_R^{\bar e\bar \ell} \sigma^j_{e f} \sigma^k_{\bar\ell\bar c}\,
\phi_{jk}(X(1)) ]\,
[\, t_L^{ef}\, \,\sigma_{cd}^g \sigma_{\bar e\bar d}^h
\,\phi_{gh}(X(0))\,]\,>\cr
&\hskip17pt + 4 \sigma_{ab}^m \,\sigma_{\bar a\bar b}^n\,
\sigma^g_{e d} \sigma^h_{\bar e \bar d}\,
\,<\phi_{mn}(X(\infty))\, [\, t_R^{\bar e\bar\ell} t_L^{e \ell}
\sigma^j_{\ell c} \sigma^k_{\bar\ell\bar c} \, \phi_{jk}(X(1)) ]\,
\phi_{gh}(X(0))>\,]\cr}}
where the $z_2,z_3$ OPE's have been evaluated in similar fashion,
the $SL(2,C)$ invariance sets $z_2\rightarrow 1$,
$z_3\rightarrow 0$, and cancellations occur among the contributions
to the OPE's from the four terms in the sum in \begads.
\vfill\eject
Then
\eqn\enads
{\eqalign{&<\,V_1(z_1,\bar z_1)\,( G^+_0 \bar G^+_0  V_2(z_2, \bar z_2))
(\tilde G^+_0 \bar{\tilde G}^+_0 V_3(z_3, \bar z_3)) \,>\cr
=& \hskip8pt 4\cdot\,[ \,{\textstyle {1\over 16}} \sigma^{mcd}
\,\sigma^{n\bar c\bar d} \,\sigma^j_{ef}\,\sigma^k_{\bar e\bar f}
\, \,<\phi_{mn}(X(\infty))\, \phi_{jk}(X(1)) \,
[\,t_L^{ef}\, t_R^{\bar e\bar f}\,\,\sigma_{cd}^g \sigma_{\bar c\bar d}^h
 \,\phi_{gh}(X(0))\,]\,>\cr
&\hskip17pt + {\textstyle {1\over 4}}
\sigma^{m cd} \,\sigma^{n\bar c \bar d}\,\, \,<\phi_{mn}(X(\infty))\,
[\, t_L^{e\ell} \sigma^j_{\ell c} \sigma^k_{\bar e\bar f} \, 
\phi_{jk}(X(1)) ]\,[\, t_R^{\bar e\bar f} \sigma^g_{ed} 
\sigma^h_{\bar c \bar d}
\,\phi_{gh}(X(0))\,]\,>\cr
&\hskip17pt + {\textstyle {1\over 4}}
\sigma^{m cd}  \,\sigma^{n\bar c\bar d} \,\, \,<\phi_{mn}(X(\infty))\,
[\, t_R^{\bar e\bar \ell} \sigma^j_{e f} \sigma^k_{\bar\ell\bar c}\,
\phi_{jk}(X(1)) ]\,[\, t_L^{ef}\, \,\sigma_{cd}^g \sigma_{\bar e\bar d}^h
\,\phi_{gh}(X(0))\,]\,>\cr
&\hskip17pt +  \sigma^{ m cd} \,\sigma^{n \bar c \bar d}\,
\sigma^g_{e d} \sigma^h_{\bar e \bar d}\,
\,<\phi_{mn}(X(\infty))\, [\, t_R^{\bar e\bar\ell} t_L^{e \ell}
\sigma^j_{\ell c} \sigma^k_{\bar\ell\bar c} \, \phi_{jk}(X(1)) ]\,
\phi_{gh}(X(0))>\,]\cr
&= 4 [\,-\bar g^{mg}\bar g^{nh}
\, \,<\phi_{mn}(x_0)\,
\phi_{jk}(x_0) \,D^j \,D^k\phi_{gh}(x_0)>\cr
&\hskip17pt + \bar g^{nh} (\bar g^{mj}\bar g^{pg}
+ \bar g^{mg}\bar g^{jp} - \bar g^{mp}\bar g^{jg} )\,
\,<\phi_{mn}(x_0)\,
D_p \phi_{jk}(x_0)
\, D^k\phi_{gh}(x_0)>\cr
&\hskip17pt + \bar g^{mg}\,
(\bar g^{nk}\bar g^{ph} + \bar g^{nh}\bar g^{kp} - \bar g^{np}\bar g^{kh} )\,
\,<\phi_{mn}(x_0)\,
D_p \phi_{jk}(x_0)
\, D^j\phi_{gh}(x_0)>\cr
&\hskip17pt - (\bar g^{mj}\bar g^{pg} + \bar g^{mg}\bar g^{jp}
- \bar g^{mp}\bar g^{jg} )\,
(\bar g^{nk}\bar g^{qh} + \bar g^{nh}\bar g^{kq} - \bar g^{nq}\bar g^{kh} )\cr
&\hskip25pt \cdot <\phi_{mn}(x_0)\,
D_p D_q \phi_{jk}(x_0)
\,\phi_{gh}(x_0)>\cr
&\hskip17pt + (\,12 \,\bar H^{mjg} \bar H^{nkh}
 - 2 \,\bar g^{mg} \bar g^{nk} \bar R^{jh}
 + 22 \,\bar g^{mg} \bar H^{pjn} \bar H_p^{\hskip3pt kh}\,)\,
\cdot \, <\phi_{mn}(x_0)\, \phi_{jk}(x_0) \phi_{gh}(x_0)>
\,]\cr}}
where the sigma matrices are expressed in terms of the
$AdS_3\times S^3$ background fields
\eqn\met{\eqalign{\bar g_{mn} &=
{\textstyle{1\over 2}}\, \sigma_m^{ab}\,\sigma_{n\,ab}\,,\quad
\bar R_{mn} \equiv - {\textstyle{1\over 2}} \,
\sigma_m^{ab} \sigma_n^{cd} \delta_{ac} \delta_{bd}\,,\quad
\bar H_{mpq} = \half\,
(\s_m \s_p \s_q)_{ab}\delta^{ab}\,,\cr}}
and the Ricci tensor is related to the
self-dual three-form flux as
$\bar R_{mn} = - \bar H_{mpq}\,\bar H_n^{\, pq}$.
Dropping terms proportional to total divergences, we have finally
from \enads
\eqn\entwoads{\eqalign{
&<\,V_1(z_1,\bar z_1)\,( G^+_0 \bar G^+_0  V_2(z_2, \bar z_2))
(\tilde G^+_0 \bar{\tilde G}^+_0 V_3(z_3, \bar z_3)) \,>\cr
&= - 12  [\,<\phi^{mn}(x_0)\,
\phi^{jk}(x_0) \,D_m \,D_n\phi_{jk}(x_0)>\,
+ 2  \, \,<\phi^{mn}(x_0)\,
D_m \phi^{jk}(x_0) \,D_j\phi_{nk}(x_0)>\cr
&\hskip30pt - 4 \,\bar H^{mjg} \bar H^{nkh} \,
<\phi_{mn}(x_0)\, \phi_{jk}(x_0) \phi_{gh}(x_0)>\cr
&\hskip30pt - 8 \, \bar g^{mg} \bar H^{pjn} \bar H_p^{\hskip3pt kh}\,
<\phi_{mn}(x_0)\, \phi_{jk}(x_0) \phi_{gh}(x_0)>\cr
&\hskip30pt -  {\textstyle{2\over 3}}
<\phi_{mn}(x_0)\, D_j \phi^{mj}(x_0) \, D_k \phi^{nk}(x_0)>\cr
&\hskip30pt  + {\textstyle{2\over 3}}\,
<\phi^{mn}(x_0)\, \phi_{mn}(x_0) \, D_j D_k \phi^{jk}(x_0)>\,]\,.\cr}}
In this derivation, we have evaluated 
terms with invariant derivatives such as 
\eqn\extra{\eqalign{&{\textstyle {1\over 16}} \sigma^{mcd}
\,\sigma^{n\bar c\bar d} \,\sigma^j_{ef}\,\sigma^k_{\bar e\bar f}
\, \,<\phi_{mn}(X(\infty))\, \phi_{jk}(X(1)) \,
[\,t_L^{ef}\, t_R^{\bar e\bar f}\,\,\sigma_{cd}^g \sigma_{\bar c\bar d}^h
 \,\phi_{gh}(X(0))\,]\,>\cr
&= -\bar g^{mg}\bar g^{nh}
\, \,<\phi_{mn}(x_0)\,
\phi_{jk}(x_0) \,D^j \,D^k\phi_{gh}(x_0)>\cr
&\hskip17pt + 2 \,\bar H^{mjg} \bar H^{nkh} \,<\phi_{mn}(x_0)\, \phi_{jk}(x_0)
\phi_{gh}(x_0)>\cr
&\hskip17pt + 2  \,\bar H^{pjn} \bar H_p^{\hskip3pt kh}
\bar g^{mg} \,<\phi_{mn}(x_0)\, \phi_{jk}(x_0)
\phi_{gh}(x_0)>\,.\cr}}

To interpret \entwoads\ on shell, we 
recall \dw\ that
the first order linearized duality equation of motion
for one of the supergravity fields related to $b_{mn}$ is 
$g^1_{njg} - {\textstyle{1\over 6}}\bar e_{njg}^{\qquad prs}\, g^1_{prs}
= - \bar H_{ngr}\, g_j^{\, r}
+ \bar H_{jgr}\, g_n^{\, r}
+ \bar H_{njr}\, g_g^{\, r}\,.$
Part of \entwoads\ is then identified as  
\eqn\gasd{\eqalign{&
\, - 4  \bar H^{mjg} \bar H^{nkh} \,
\,g_{mn}\, g_{jk}\, g_{gh}
- 8 \,  \bar g^{mg} \bar H^{pjn} \bar H_p^{\hskip3pt kh}\,
\,g_{mn}\, g_{jk}\,g_{gh}\cr
&= - 4 \bar H^{nkh} \, g _{jk} \, g_{gh}\,
( g^{1\hskip5pt jg}_n - {\textstyle{1\over 6}} \bar e_n^{\hskip3pt jgrst}
\,g^1_{rst}\,)\,. \cr}}
Assuming that the only non-vanishing string field fluctuation is $g_{mn}$,
we have $\phi_{mn} = g_{mn}$ and
the gauge condition becomes $D^m g_{mn} = 0$.

Then on $AdS_3\times S^3$, the string theory three graviton amplitude is
\eqn\adstga{
\eqalign{&<\,V_1(z_1,\bar z_1)\,( G^+_0 \bar G^+_0  V_2(z_2, \bar z_2))
(\tilde G^+_0 \bar{\tilde G}^+_0 V_3(z_3, \bar z_3)) \,>\cr
=&- 12  [\,<g^{mn}(x_0)\,
g^{jk}(x_0) \,D_m \,D_n g_{jk}(x_0)>\,
+ 2  \, \,< g^{mn}(x_0)\,
D_m g^{jk}(x_0) \,D_j g_{nk}(x_0)>\cr
&\hskip30pt - 4 \,<\, \bar H^{mjg} (x_0) \bar H^{nkh}(x_0) \,
g_{mn}(x_0)\, g_{jk}(x_0) g_{gh}(x_0)>\cr
&\hskip30pt - 8 \, <\,\bar g^{mg}(x_0) \bar H^{pjn}(x_0) 
\bar H_p^{\hskip3pt kh}(x_0)\,
g_{mn}(x_0)\, g_{jk}(x_0) g_{gh}(x_0)>\,]\cr
&= -12\, \int \,d^6 x {\sqrt{\bar g}} \,
[g^{mn}\,g^{jk}\, D_m \,D_n g_{jk}\,
+2 \,\, g^{mn}\,D_m g^{jk}\, D_j g_{nk}\,]\,,\cr}}
where $D_m$ is the covariant derivative on 
$AdS_3\times S^3$. \adstga\ is the curved space analog of
\flat. 
In three dimensions, the graviton has no propagating
degrees of freedom, which means the graviton field can be gauged to
zero. Nonetheless, expanding \adstga\ in
spherical harmonics on $S^3$, and noting that $\bar g_{mn} = \bar g_{\mu\nu},\,
\bar g_{\alpha\beta}$, the
background metric of $AdS_3\times S^3$ for $1\le \mu,\nu\le 3$ and
$1\le \alpha,\beta\le 3$, we find that the covariant derivatives
factorize and 
the string theory calculation retains the familiar structure 
$\sim \int d^3x  \,{\sqrt{\bar g_3}} \,
[g^{\mu\nu}\,g^{\rho\sigma}\, D_\mu \,D_\nu g_{\rho\sigma}\,
+2 \,\, g^{\mu\nu}\,D_\mu g^{\rho\sigma}\, D_\rho g_{\nu\sigma}\,]\,.$

In general, $\alpha'$ corrections are expected to occur in
four or higher $n$-point string tree amplitudes, but 
will be calculable only as an expansion in $\alpha'$ since
the worldsheet theory is not free. 
To study the AdS/CFT correspondence,
the bulk correlations functions on shell can be related to 
correlations on the boundary. Since $\alpha'$ is related
the coupling constant of the spacetime conformal field theory,
to investigate this correspondence systematically
it would be of interest to attain tree-level expressions 
that are exact in $\alpha'$, 
perhaps by adapting integrable methods for
sigma models which have a supergroup manifold target space \rs\ such as
this $AdS_3\times S^3$ theory.

\vskip50pt
{\bf Acknowledgements:}
We thank Nathan Berkovits and Edward Witten for conversations.
LD and KB are partially supported by the U.S. Department of Energy,
Grant No. DE-FG02-97ER-41036/Task A.

\listrefs

\bye